\documentclass[11pt,onecolumn,a4paper]{article}
\usepackage[T1]{fontenc}
\usepackage{newtxtext,newtxmath}
\usepackage[margin=25mm,footskip=11mm]{geometry}
\usepackage{amsmath}
\usepackage{graphicx,microtype,needspace,placeins}
\usepackage[numbers,sort&compress]{natbib}
\usepackage[colorlinks=true,allcolors=blue]{hyperref}
\usepackage{titlesec}
\usepackage[font=small,labelfont=bf]{caption}
\titleformat{\section}{\normalfont\large\bfseries}{\thesection.}{0.6em}{}
\titleformat{\subsection}{\normalfont\normalsize\bfseries}{\thesubsection.}{0.6em}{}
\titlespacing*{\section}{0pt}{16pt plus 2pt minus 2pt}{7pt}
\titlespacing*{\subsection}{0pt}{13pt plus 2pt minus 1pt}{6pt}
\newcommand{\GeV}{\,\mathrm{GeV}}

\hypersetup{pdftitle={A Common Antimatter Response in AMS-02 Positrons and Antiprotons},pdfauthor={Yi Yang},pdfsubject={Preprint v29: time orientation, cosmic-ray history overlap, and AMS spectral tests}}
\begin{document}
\begin{center}
{\fontsize{17}{21}\selectfont\bfseries A Common Antimatter Response\\
in AMS-02 Positrons and Antiprotons\par}
\vspace{12pt}
{\large Yi Yang$^{1,2}$\par}
\vspace{6pt}
{\small $^1$Institute of Physics, Academia Sinica, Taipei 11529, Taiwan\par
$^2$Department of Physics, National Cheng Kung University, Tainan 70101, Taiwan\par
\href{mailto:yiyang429@as.edu.tw}{yiyang429@as.edu.tw}\par}
\end{center}
\vspace{2pt}
\begin{abstract}
I develop a common history-overlap description that gives antiparticle time orientation a physical role beyond its use in field-theory amplitudes. In this framework, opposite orientation reduces the effective overlap with a finite, matter-defined environmental history, while standard local interactions, positive physical energies, and causal detector records are retained. The common geometry connects two contrasting cosmic-ray spectral patterns measured by AMS-02. For positrons, a geometric exposure estimate displaces the broad maximum of an empirical energy-weighted electron reference from 8--10 GeV to the 300--400 GeV region. A number-conserving calculation combines energy redistribution and spatial dilution to interpret the relative weighted height, of order $0.1$. For antiprotons, ordinary secondary production retains the parent-proton history. I extend the same three-direction overlap estimate to the accumulated logarithmic deformation after production. Within this approximation, an ordinary softening of a few tenths leaves a residual additional spectral index of order 0.01, yielding near preservation of the source index and a nearly flat antiproton-to-proton ratio in the production-scaling limit. The positron comparison uses no additional dominant source, and known astrophysical contributions remain allowed. The paired AMS spectra give this time-orientation framework concrete empirical direction: a common geometric description relates a lepton scale displacement to hadronic index preservation and defines explicit targets for independent theoretical and experimental tests.
\end{abstract}
\par
\section{Introduction}
\label{sec:intro}

The Stueckelberg--Feynman description represents a positron by an oppositely time-oriented electron line \cite{Stueckelberg,Feynman}. Its role in field-theory amplitudes is well established \cite{Peskin}. The central proposal of this work is that relative time orientation also organizes the environmental histories contributing to an observed antiparticle population. Following the time-orientation formulation of quantum field theory in Ref.~\cite{YangTime}, I distinguish phase orientation relative to matter clocks from the sequence of positive-energy detector events. I develop the accumulated-response consequences of this interpretation in a finite matter environment, retaining standard local QED and QCD.

Galactic cosmic rays provide the macroscopic setting for this question. Their spectra combine the history of injection or secondary production with the subsequent effects of Galactic magnetic fields, radiation backgrounds, and interstellar matter \cite{StrongReview}. The observables therefore carry information about accumulated environmental response as well as local production. The Alpha Magnetic Spectrometer (AMS-02) measures cosmic-ray electrons, positrons, protons, and antiprotons separately \cite{AMSpositron,AMSelectron,AMSbarp2016}. These charge-resolved spectra make the proposed connection between time orientation and environmental history accessible to observational comparison.

The AMS measurements show two complementary patterns. The cosmic-ray positron spectrum departs from its low-energy trend above about 25 GeV and develops a dropoff in the few-hundred-GeV region, while the electron spectrum has a different high-energy structure \cite{AMSpositron,AMSelectron}. Above roughly 60 GV, antiprotons have nearly the same rigidity dependence as protons, giving an approximately constant antiproton-to-proton ratio \cite{AMSbarp2016}. The proposed response connects these observations through a displacement of the lepton reference scale and approximate preservation of the secondary antiproton index.

In the history-overlap picture developed here, the oppositely oriented sector has a reduced effective overlap with the finite history of the matter-defined environment. For cosmic-ray leptons, the same local radiative-loss law acts over different effective exposures, producing different accumulated cooling. For secondary antiprotons, the same geometric estimate is extended to the logarithmic spectral deformation accumulated after production, leaving a small residual softening. The unifying idea is a geometric reduction of accumulated response: the local cooling and collision processes retain their standard form, while the contributing histories differ.

The lepton normalization illustration uses equal injected differential particle counts for corresponding electron and positron ensembles. This component is distinct from the full measured electron population, which includes ordinary primary electrons. Known pulsars can also supply energetic pairs \cite{HooperPulsars}. The central comparison treats their source-specific contributions as subleading and describes the main positron feature through the history response; it does not determine their fractions.

Geometric estimates give the time-orientation interpretation explicit spectral consequences in characteristic energies, relative heights, and leading indices, establishing a focused direction for theoretical and experimental tests. A related cosmological application of history geometry was discussed in Ref.~\cite{YangYeung}; the present analysis concerns Galactic cosmic rays and uses no cosmological expansion factor. The positron response was explored separately in Ref.~\cite{YangPositron}; the present work develops a common geometric description of the lepton and antiproton channels.

Section~\ref{sec:framework} defines the common history-overlap response. Sections~\ref{sec:exposure}--\ref{sec:lepton-data} develop the positron scale shift, relative-height interpretation, and data comparison in that order. Section~\ref{sec:antiproton} retains ordinary secondary production and uses the geometric overlap to estimate the residual post-production softening before comparing with the antiproton plateau.

\section{Time orientation and the common history-overlap response}
\label{sec:framework}

Particle and antiparticle modes appear in the standard Dirac-field expansion with phases $e^{-ip\cdot x}$ and $e^{+ip\cdot x}$, respectively, with $p^0>0$. Here $x$ is the spacetime coordinate. On these modes, the eigenvalue $K^\mu$ of $i\partial^\mu$ is $+p^\mu$ or $-p^\mu$. The time-orientation formulation distinguishes this phase momentum from the positive-energy physical momentum \cite{YangTime}:
\begin{equation}
s_T=\operatorname{sign}(K^0),\qquad p^\mu=s_TK^\mu,\qquad p^0>0.
\label{eq:orientation}
\end{equation}
The label $s_T$ records phase orientation relative to matter time, while both members of a produced pair generate outgoing tracks in increasing laboratory time. Local QED amplitudes and detector responses retain their usual interpretation.\footnote{The standard retarded response concerns the full Dirac field, including both charge sectors; no separate advanced causal propagator is assigned to positrons \cite{Peskin}.} The same orientation convention is used for an asymptotic antiproton, whose production and composite structure remain governed by ordinary QCD.

Relative time orientation provides the conceptual foundation of this history-overlap description. An observed cosmic-ray population receives contributions from source events at different positions and times, together with the fields and matter encountered between production and detection. This source and environmental history has finite spatial and temporal support. I use \emph{finite-history overlap} for the effective contribution of those histories to the accumulated population response. Within the framework, opposite orientation reduces that overlap. The physical proposal is therefore an orientation dependence of environmental sampling, beyond the finite extent already present in an ordinary Galactic description.

For leptons, accumulated radiative exposure measures the sampled radiation and magnetic background. The local loss equation evolves the energy for a specified exposure, and the history-overlap approximation relates the effective exposures of the two charge sectors. The history picture and the exposure relation are thus the physical description and its quantitative implementation. For antiprotons, secondary collisions supply the source, and the accumulated response connects that production to the present intensity. The parent-proton history already contained in the source and the antiproton history after production remain distinct.

To give this picture a common spectral realization, I retain a fixed horizontal scale and a nearly constant amplitude at leading order. In the following notation, $x$ denotes energy or positive rigidity magnitude, rather than the spacetime coordinate above. Let $B(x)$ be a reference spectrum, $\xi$ a dimensionless horizontal factor, and $\mathcal A$ a dimensionless relative amplitude:
\begin{equation}
B(x)\longrightarrow\mathcal A B(\xi x),\qquad 0<\xi\leq1.
\label{eq:response}
\end{equation}
A constant $\mathcal A$ changes the height without adding a spectral power; this leading limit is called \emph{scale neutrality}, $d\ln\mathcal A/d\ln x\simeq0$. The small residual change of spectral index is estimated in Sec.~\ref{sec:pbar-history} by extending the three-direction history overlap of Sec.~\ref{sec:exposure} to the accumulated logarithmic deformation. This shared weighting is the effective geometric approximation that connects the two channels.

In the constant-amplitude limit, two consequences follow directly. Define $x_*$ as the position of a specified reference feature, such as a maximum of a stated energy-weighted spectrum. The rescaled argument reaches that feature when $\xi x=x_*$, so its position moves to $x_*/\xi$; a constant amplitude does not move it. For a pure power law $B(x)\propto x^{-\gamma}$ with spectral index $\gamma$, $B(\xi x)=\xi^{-\gamma}B(x)$: the fixed dilation changes only the normalization. A displaced feature and a preserved index are the two leading manifestations of the common response. Section~\ref{sec:pbar-history} then estimates the small remaining change of index from the history geometry.

The physical reference is specified in each application. The lepton comparison uses the measured electron spectrum as an empirical ordinary-history reference, with an exposure relation and a number-conserving normalization illustration. The antiproton comparison uses the secondary-production spectrum as the input to the \emph{total post-production} response; the dimensional conversion from source rate to intensity is kept explicit. The two channels retain their own production physics and normalizations. What they share is the geometric reduction of the effective history contributing to the accumulated response.

\section{Positrons: a displaced spectral scale}
\label{sec:positron}
\subsection{Effective exposure and the energy-scale shift}
\label{sec:exposure}

Electrons and positrons obey the same leading radiative energy-loss law. Writing $t_M$ for matter-clock time, a useful approximation is \cite{BlumenthalGould}
\begin{equation}
\frac{dE}{dt_M}=-\beta u E^2,\qquad
\frac{1}{E}-\frac{1}{E_{\rm inj}}=\beta\mathcal X.
\label{eq:loss}
\end{equation}
Here $E_{\rm inj}$ is initial energy, $u$ the effective radiation and magnetic energy density, and $\beta>0$ the loss coefficient. The accumulated exposure $\mathcal X=\int u\,dt_M$ is evaluated along the effective history defined in Sec.~\ref{sec:framework}. The inverse-Compton part uses the Thomson-limit scaling. Both charges lose physical energy with the same sign; a smaller $\mathcal X$ means less accumulated cooling. Equation~\eqref{eq:loss} gives the characteristic radiative scale $(\beta\mathcal X)^{-1}$, which increases when the effective exposure decreases.

Write $\mathcal X_+$ for the positron exposure and $\mathcal X_-$ for the electron exposure. The three-dimensional overlap approximation factorizes their relative history support into three directions. The factor $q_i$, with $i=x,y,z$, is the positron-to-electron effective history-reach ratio along that direction:
\begin{equation}
\xi_G\equiv\frac{\mathcal X_+}{\mathcal X_-}
\simeq q_xq_yq_z\equiv q^3.
\label{eq:overlap}
\end{equation}
The shorthand $q$ is the geometric mean of the three factors. This factorized overlap approximates the relative sampling of the background in the exposure integral; it does not equate a path integral with an ordinary volume.

The order-of-magnitude estimate is guided by the hierarchy of Galactic vertical and radial scales \cite{StrongMoskalenko,Evoli}. A halo half-height of order 5 kpc and an illustrative radial source extent of 15--20 kpc give a ratio of about 0.25--0.33. This suggests the rounded effective estimate $q\simeq0.30$. The vertical scale refers to the cosmic-ray halo, not the stellar or gas thin disk. Secondary-nuclei analyses permit a halo of this order or larger under their cross-section assumptions \cite{Evoli}. The estimate uses this scale hierarchy as a guide to the overlap, rather than as a measurement of a lepton's history reach.

Anisotropy is retained through $q_xq_yq_z$. Similarly flattened history domains can have a product of three scale ratios just as spherical domains can; only a restriction acting in one direction alone would supply a single factor. The Galactic aspect ratio and the ratios between the two sector histories are different quantities. Using the former to estimate the geometric mean of the latter is the modeling step. The directional factors themselves, and any compensation among them, are not fixed by this approximation.

I use $q=0.30$, hence $\xi_G=0.027$, throughout the spectral comparison. Equation~\eqref{eq:loss} then raises the characteristic radiative scale by about 37. To connect this scale to a measured spectral feature, the approximately shape-preserving response of Eq.~\eqref{eq:response} is applied to an empirical electron reference.

For the lepton spectra, write $\Phi_i(E)$ for the differential intensity of species $i$ and $\mathcal S_i(E)=E^3\Phi_i(E)$ for its energy-weighted form. The weight makes broad spectral structures visible; a peak in this comparison means a maximum of $\mathcal S_i$, not of the unweighted flux. The measured low-energy electron flux supplies the reference $\Phi_{e,\rm ref}$, with weighted form $\mathcal S_{e,\rm ref}$. Its broad maximum in the 4--20 GeV reference interval lies near 8--10 GeV \cite{AMSelectron}. The reference provides the ordinary-history response shape: it includes primary-electron and heliospheric contributions and is not an isolated pair-source spectrum or a measurement of an intrinsic interstellar cooling scale.

Let $E_{e,*}$ denote this reference maximum and $E_{+,*}$ its counterpart in the mapped positron spectrum; the asterisk labels the stated spectral feature. In the single-scale description, the reference argument changes from $E$ to $\xi_GE$. The same feature is therefore reached when $\xi_GE=E_{e,*}$, giving
\begin{equation}
E_{+,*}=\frac{E_{e,*}}{\xi_G}\sim300\text{--}400\GeV.
\label{eq:peak}
\end{equation}
Thus the geometric exposure estimate maps the electron reference's few-GeV structure into the few-hundred-GeV positron region, independently of the constant amplitude. These are broad spectral regions, not fitted confidence intervals. The displaced maximum lies in the region of the AMS positron dropoff \cite{AMSpositron}; an $E^3\Phi$ maximum, a spectral break, and an exponential cutoff are distinct quantities. This completes the energy-scale argument. The next question is the height of the displaced structure.

\subsection{Geometric dilution and the relative normalization}
\label{sec:normalization}

The position of the mapped maximum is now specified, but its relative height is not. To interpret the intensity difference, I combine particle-number conservation under the energy transformation with spatial dilution. Reduced exposure changes how particles populate energy intervals; distributing the same particle content over a larger effective region lowers their local density. These two effects determine the relative amplitude in the illustration below without adding another positron population.

Let $a>1$ be the dimensionless ratio of the positron profile width to that of its electron counterpart in this illustration. For similarly shaped, normalized three-dimensional profiles rescaled by this factor along each direction, the central density scales as $a^{-3}$ while the total particle number is unchanged. The same volume factor applies to similarly flattened profiles. For an observer in the nearly uniform central region, the local intensity receives this factor because intensity is proportional to local density for a nearly isotropic ultrarelativistic population \cite{PDGCosmic}.

To combine this dilution with the energy rescaling, take two finite electron and positron ensembles with the same injected differential count $\nu_{\rm inj}(E_{\rm inj})\propto E_{\rm inj}^{-\Gamma}$, including equal normalization over a common finite energy interval. Here $\Gamma$ is the injection index and $\nu_s(E)=dN_s/dE$ is the count spectrum after exposure, with $s=+,-$ labeling charge. For this illustration, the spatial profile is taken to be approximately energy independent over the comparison interval, and each ensemble is assigned a fixed $\mathcal X_s$. Equation~\eqref{eq:loss} then gives $E_{\rm inj}=E/(1-\beta\mathcal X_sE)$, within the range $1-\beta\mathcal X_sE>0$. Conservation in corresponding energy bins, $\nu_s(E)\,dE=\nu_{\rm inj}(E_{\rm inj})\,dE_{\rm inj}$, yields
\[
\begin{aligned}
\nu_s(E)&=\nu_{\rm inj}\!\left(\frac{E}{1-\beta\mathcal X_sE}\right)
          \frac{1}{(1-\beta\mathcal X_sE)^2}\\
&\propto E^{-\Gamma}(1-\beta\mathcal X_sE)^{\Gamma-2}.
\end{aligned}
\]
Since $\mathcal X_+=\xi_G\mathcal X_-$, the last factor is identical for $\nu_+(E)$ and $\nu_-(\xi_GE)$, giving $\nu_+(E)=\xi_G^\Gamma\nu_-(\xi_GE)$. Write $\Phi_\pm$ for the local intensities of these two illustrative ensembles. Including the spatial dilution therefore gives
\begin{equation}
\Phi_+(E)\simeq a^{-3}\xi_G^\Gamma\Phi_-(\xi_GE).
\label{eq:dilution}
\end{equation}
Both corresponding initial energies must lie in the common injection interval. This relation compares the electron and positron distributions at different final energies, $\xi_GE$ and $E$, which also correspond to different initial energies. It does not multiply the energy of each particle from the same injection event by a fixed factor. Each ensemble conserves its count over its own complete cooled interval. The factor $\xi_G^\Gamma$ follows from the common injection spectrum and the energy-bin Jacobian. This illustration supplies a normalization relation, not an independent derivation of the measured electron reference shape.

In the weighted representation defined in Sec.~\ref{sec:exposure}, Eq.~\eqref{eq:dilution} becomes $\mathcal S_+(E)\simeq a^{-3}\xi_G^{\Gamma-3}\mathcal S_-(\xi_GE)$. The additional factor $\xi_G^{-3}$ comes from comparing the weights $E^3$ and $(\xi_GE)^3$, rather than from an extra physical dilution. The relative weighted amplitude is therefore $a^{-3}\xi_G^{\Gamma-3}$. For a representative injection index $\Gamma=2.5$ and the fixed exposure ratio $\xi_G=0.027$, the energy-redistribution factor alone is $\xi_G^{-1/2}\simeq6.1$.

The observed relative height of the broad spectral structures motivates a weighted amplitude of order $0.1$ at the corresponding energies $E$ and $\xi_GE$. In this number-conserving illustration, $a^{-3}\times6.1\sim0.1$ therefore requires $a\sim(6.1/0.1)^{1/3}\simeq3.9$. I use $a\simeq4$ as the representative broadening: the differential-flux coefficient is then about $1.9\times10^{-6}$ and the weighted amplitude is $0.095\simeq0.1$. The rounded value $0.1$ defines the central spectral comparison.

The two geometric quantities have distinct roles: $q$ estimates the exposure ratio through $\xi_G=q^3$, whereas $a$ specifies the spatial broadening inferred from the relative height at that exposure scale. Thus $a\simeq4$ is an interpretation of the known normalization difference within the equal-injection, rescaled-profile illustration. At fixed $\Gamma=2.5$, taking $a=3$--$5$ gives differential-flux coefficients $(0.96$--$4.4)\times10^{-6}$, or weighted amplitudes $0.049$--$0.225$. This range illustrates the sensitivity to factor-of-a-few broadening; it is not an independently measured spatial size or a confidence interval.

\subsection{Comparison with the AMS-02 spectra}
\label{sec:lepton-data}

\begin{figure}[htbp]
\centering
\includegraphics[width=0.98\textwidth]{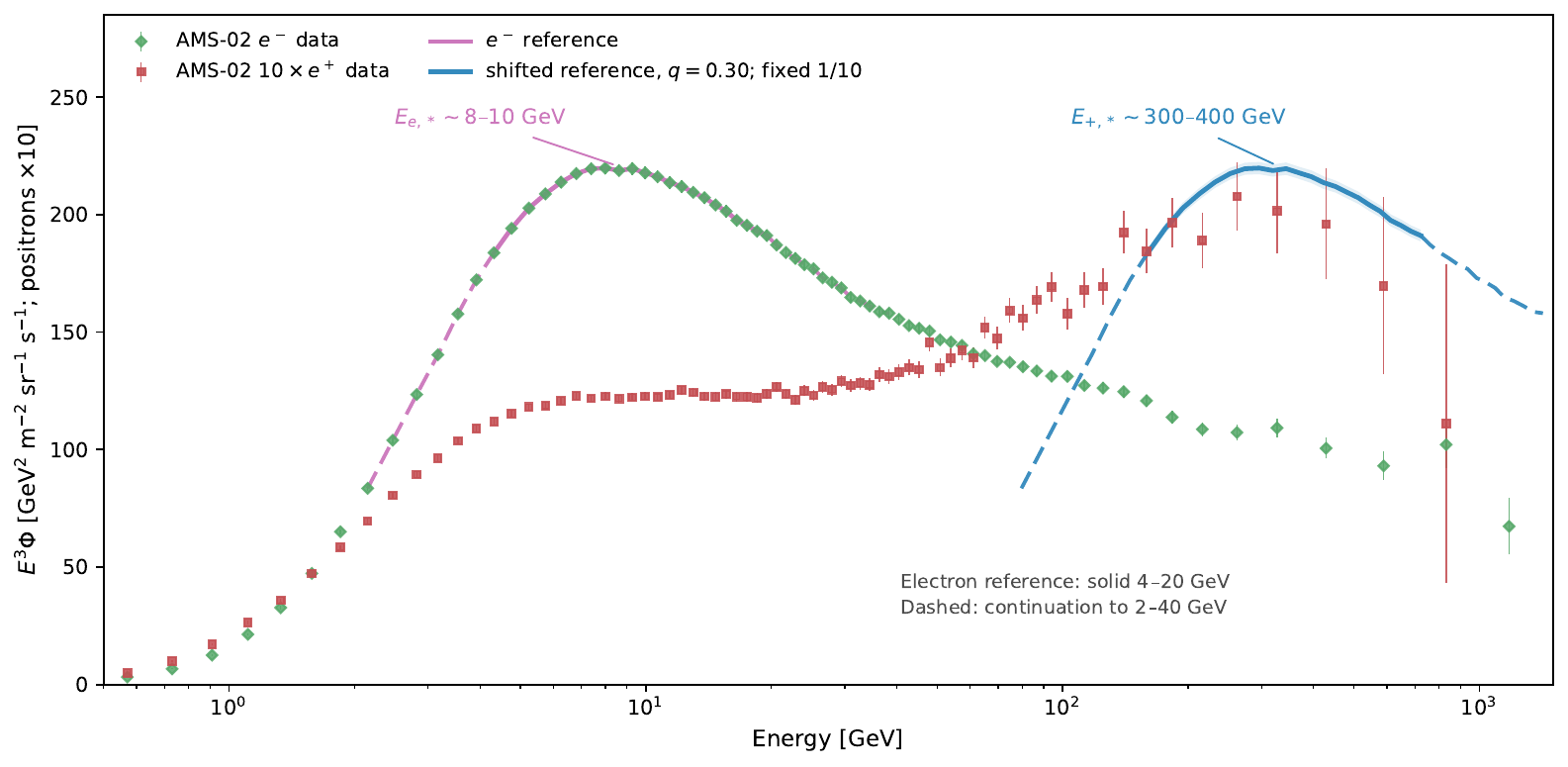}
\caption{AMS-02 electron data and positron data multiplied by ten, displayed as $E^3\Phi$ on a logarithmic energy axis \cite{AMSpositron,AMSelectron}. Purple curves join the measured electron reference; the blue curve uses $q=0.30$ and the rounded weighted amplitude $0.1$ in Eq.~\eqref{eq:positron-map}. Solid segments use 22 electron points in the 4--20 GeV selection, with mapped representative energies of 160--723 GeV. The narrow pale-blue strip carries only the electron-reference errors at the fixed central amplitude, not a geometric or model uncertainty. Dashed continuations retain the same mapping and use adjacent measured electron entries from 2 to 40 GeV, reaching about 80--1420 GeV. They explore the mapping outside the central interval and extend beyond the 1 TeV upper edge of the positron data. Labels indicate broad maximum regions.}
\label{fig:positron}
\end{figure}

The energy displacement in Eq.~\eqref{eq:peak} and the amplitude interpretation of Sec.~\ref{sec:normalization} now specify the central comparison. Applying both to the empirical electron reference defined in Sec.~\ref{sec:exposure} gives
\begin{equation}
\begin{aligned}
\mathcal S_+(E)&\simeq\frac{1}{10}\mathcal S_{e,\rm ref}(\xi_GE),\\
\Phi_+(E)&\simeq\frac{\xi_G^3}{10}\Phi_{e,\rm ref}(\xi_GE).
\end{aligned}
\label{eq:positron-map}
\end{equation}
The second line follows by removing the $E^3$ weight at the corresponding energies; its differential-flux coefficient is $\xi_G^3/10\simeq2\times10^{-6}$. The factor ten compares weighted spectra at $E$ and $\xi_GE$, not the positron and electron fluxes at one energy. The positron data and the mapped curve are both multiplied by ten in Fig.~\ref{fig:positron}, so their broad structures can be compared directly with the electron reference on the same vertical scale.

For the plotted reference, take $\mathcal S_{e,\rm ref}$ directly from the 22 tabulated electron points with representative energies between 4 and 20 GeV \cite{AMSelectron}. Line segments join successive values on a logarithmic energy axis, retaining the measured local shape without a fitted spectral function. These points form the solid reference and its solid mapped counterpart. Dashed continuations use adjacent electron data selected from 2 to 40 GeV, extending both selection limits by a factor of two. No horizontal scale or amplitude is readjusted in the continuation. The narrow pale-blue strip propagates only the electron-reference errors at the fixed amplitude $0.1$; it is not a geometric or model uncertainty band.

All published electron and positron points are retained for context. The 41 reference energies span 2.15--38.33 GeV, mapping to approximately 80--1420 GeV; statistical and total systematic errors are combined in quadrature. The displaced structure has a similar scale and height to the central high-energy positron data. The comparison tests these orders of magnitude, rather than bin-level agreement, curvature significance, or source fractions.

\Needspace{5\baselineskip}
The dashed central continuation falls below the positron data toward 80--120 GeV and declines more slowly than the highest-energy positron central values. These differences remain visible because the same horizontal scale and amplitude are retained in the continuation. The empirical reference and the single-scale approximation delimit the part of the positron spectrum addressed here.

\section{Antiprotons: approximate spectral-index preservation}
\label{sec:antiproton}

The positron reference reveals the common response through a displaced maximum. For an approximately power-law antiproton source, the corresponding response quantity is the additional change of index after production. The argument retains the parent-proton history and local secondary collision, identifies the ordinary additional softening, and estimates its residual size using the same geometric overlap. The resulting small, nonzero index estimates the response contribution to departures from the leading scale-neutral description.

\subsection{Secondary production: the source spectrum}
\label{sec:pbar-production}

High-energy protons and heavier nuclei produce secondary antiprotons in collisions with interstellar gas. Let $R=|pc/(Ze)|$ denote positive rigidity magnitude, with $p$ the momentum magnitude, $Z$ the charge number, and $e$ the elementary charge. Define $Q_{\bar p}(R)$ as the number of antiprotons produced per unit volume, time, and rigidity, with units $\mathrm{m}^{-3}\,\mathrm{s}^{-1}\,\mathrm{GV}^{-1}$. The expression below gives the proton-on-hydrogen contribution; other nuclear channels can be included similarly \cite{Winkler,Korsmeier}.

Let $n_{\rm H}$ be the hydrogen number density and $\Phi_p(R_p)$ the isotropic incident proton intensity per unit parent rigidity $R_p$. Its units are $\mathrm{m}^{-2}\,\mathrm{s}^{-1}\,\mathrm{sr}^{-1}\,\mathrm{GV}^{-1}$. In the representative production region, approximate its shape by the local proton reference, $\Phi_p(R)\propto R^{-\gamma_p}$, where $\gamma_p$ is the proton spectral index. This approximation must cover the contributing parent rigidities, not only the observed antiproton interval.

In the ideal high-energy scaling limit, take $d\sigma/dR=(\sigma_0/R_p)k(R/R_p)$, where $\sigma_0$ is a constant cross-section scale and $k$ is a dimensionless inclusive distribution in the daughter-to-parent rigidity ratio. Neglecting threshold scales, the source is
\begin{equation}
\begin{aligned}
Q_{\bar p}(R)
&=4\pi n_{\rm H}\int_R^\infty dR_p\,\Phi_p(R_p)
\frac{d\sigma}{dR}(R_p,R)\\
&\simeq4\pi n_{\rm H}\sigma_0\int_R^\infty\frac{dR_p}{R_p}\,
\Phi_p(R_p)\,k(R/R_p)
\propto R^{-\gamma_p}.
\end{aligned}
\label{eq:pbar-source}
\end{equation}
The factor $4\pi$ integrates the incident intensity over solid angle, so the expression has the units of a production rate, not an observed intensity. The index is preserved because the inclusive distribution depends on a rigidity ratio: summing over the higher-rigidity parents leaves a constant moment multiplying the proton spectrum.\footnote{With the integration variable $z=R/R_p$, the integral in the second line is $\Phi_p(R)\int_0^1 dz\,z^{\gamma_p-1}k(z)$. The moment must converge and the parent spectrum must obey the stated scaling. Here $z$ is only the daughter-to-parent fraction used in the production integral, not a history-overlap parameter.} The resulting $Q_{\bar p}(R)\propto\Phi_p(R)$ fixes the source shape. A further step is needed to obtain the intensity at the observer.

\subsection{The two stages of residence and accumulation}
\label{sec:two-stages}
The incident proton spectrum in Eq.~\eqref{eq:pbar-source} is already the result of source injection and Galactic propagation. Its confinement history affects how many protons are available to collide with gas, and is retained when that measured spectrum is used to construct $Q_{\bar p}$. Reducing the parent matter column would reduce antiproton production itself, which is not the effect proposed here. Nor should an additional parent residence factor be multiplied onto the already propagated proton intensity.

Once a collision produces an antiproton, a second population history begins. Antiprotons made at different times accumulate in the observable population until escape or removal. In a steady, spatially averaged, escape-dominated illustration, a constant production rate builds a number density equal to that rate times the mean residence time. Let $\tau_{\rm esc}(R)$ denote this ordinary post-production residence time and $\Phi_{\bar p}^{\rm esc}(R)$ the corresponding intensity. Neglecting energy redistribution and destruction in this illustration gives \cite{StrongReview}
\begin{equation}
\Phi_{\bar p}^{\rm esc}(R)\simeq
\frac{c}{4\pi}\,Q_{\bar p}(R)\,\tau_{\rm esc}(R).
\label{eq:pbar-escape}
\end{equation}
Here $c$ is the ultrarelativistic particle speed. The product $Q_{\bar p}\tau_{\rm esc}$ is a differential number density, converted to intensity by $c/(4\pi)$ as in Sec.~\ref{sec:normalization}.

In the illustrative scaling $\tau_{\rm esc}(R)\propto R^{-\delta}$, with $\delta>0$, the combination with $Q_{\bar p}(R)\propto\Phi_p(R)$ gives $\Phi_{\bar p}^{\rm esc}/\Phi_p\propto R^{-\delta}$. Lower-rigidity particles accumulate for longer, so this second stage softens the secondary spectrum relative to its parent. The first residence history is already contained in $\Phi_p$ and hence in the source; the second factor appears after secondary production. They belong to two different populations, not to two passages of the same antiproton. This two-stage accounting identifies where the proposed history dependence enters. Production kinematics, cross sections, and spatial evolution contribute to a more detailed description of the ratio.

\subsection{Geometric estimate of the residual softening}
\label{sec:pbar-history}

The parent proton belongs to the matter-oriented sector; the produced antiproton has the opposite phase orientation of Eq.~\eqref{eq:orientation}. The history-overlap description assigns this second population a reduced effective overlap with the matter-defined environment. The source spectrum and the first population history remain unchanged. The question is not whether antiprotons escape, but how much rigidity-dependent deformation the total post-production response retains.

Let $C_{\bar p}(R)$ be the source-to-intensity coefficient in $\Phi_{\bar p}(R)\simeq C_{\bar p}(R)Q_{\bar p}(R)$. Its units are length per solid angle, and it includes spatial reach, survival, and accumulation. Its rigidity dependence is the quantity estimated here. In the ordinary reference of Eq.~\eqref{eq:pbar-escape}, the coefficient is $c\tau_{\rm esc}(R)/(4\pi)$ and minus its logarithmic slope supplies the positive softening index $\delta$. The geometric estimate therefore addresses the rigidity-dependent change of spectral shape, not the absolute response normalization.

For the antiproton channel, I extend the three-direction overlap of Eq.~\eqref{eq:overlap} to the accumulated \emph{logarithmic} deformation. The geometric approximation treats the coarse-grained logarithmic response as additive over its contributing history, with the retained support representative of the spectral-deforming contributions. Under this representative sampling, the retained logarithmic change is of order $\xi_G$ times that of the ordinary reference. The common factor thus weights radiative exposure in the lepton channel and logarithmic spectral deformation in the antiproton channel.

Choose a fixed reference rigidity $R_0$ to remove the arbitrary normalization. For nearly rigidity-independent overlap and smooth power-law behavior, take the positive proportionality factor to be of order unity and approximately constant across the comparison interval. The normalized response and its residual index then obey the estimates
\begin{equation}
\begin{aligned}
\ln\frac{C_{\bar p}(R)}{C_{\bar p}(R_0)}
&\sim\xi_G\ln\frac{\tau_{\rm esc}(R)}{\tau_{\rm esc}(R_0)},\\
\delta_{\rm eff}\equiv-\frac{d\ln C_{\bar p}}{d\ln R}
&\sim\xi_G\delta\sim(q_xq_yq_z)\delta.
\end{aligned}
\label{eq:pbar-geometric-softening}
\end{equation}
Both logarithms act on dimensionless ratios. The index $\delta_{\rm eff}$ characterizes the remaining deformation of the total antiproton response, while $\delta$ characterizes the ordinary reference. Equation~\eqref{eq:pbar-geometric-softening} is the shared-overlap estimate, with an order-one coefficient reflecting the different environmental weights of radiative exposure and spectral deformation. Its validity requires the representative sampling stated above: a boundary-dominated or otherwise strongly concentrated deformation need not follow this scaling.

A constant shortening $\tau_{\rm eff}(R)=\xi_G\tau_{\rm esc}(R)$ would leave the extra index $\delta$ unchanged. Equation~\eqref{eq:pbar-geometric-softening} instead reduces the \emph{normalized logarithmic variation} of the total post-production response. It replaces the ordinary reference at this stage; the two residence factors are not multiplied together. A fixed dilation of a power law is absorbed into normalization, so no additional antiproton horizontal-scale parameter is needed.

Using the common geometric estimate $\xi_G=0.027$, an illustrative ordinary softening $\delta\sim0.3$--$0.5$ gives $\delta_{\rm eff}\sim10^{-2}$. Taking the proportionality coefficient as unity for a representative estimate, the response changes by about 2--3\% from 60 to 500 GV. This is a geometric order estimate obtained without fitting the antiproton plateau. It explains why approximate index preservation is the leading outcome of the construction, while retaining a small residual tilt.

Treating $\delta_{\rm eff}$ as approximately constant over this interval and combining the residual response with the source-scaling result of Eq.~\eqref{eq:pbar-source} gives
\begin{equation}
\begin{aligned}
\Phi_{\bar p}(R)&\simeq C_{\bar p}(R)\,Q_{\bar p}(R),\\
\frac{\Phi_{\bar p}(R)}{\Phi_p(R)}
&\propto\left(\frac{R}{R_0}\right)^{-\delta_{\rm eff}},
\qquad \delta_{\rm eff}\sim\xi_G\delta.
\end{aligned}
\label{eq:pbar-scaling}
\end{equation}
The small remaining index gives approximate preservation of the source slope. A horizontal ratio is the leading representation when $|\delta_{\rm eff}\ln(R/R_0)|\ll1$. The same geometric fraction therefore relates a displaced radiative scale for positrons to a small residual logarithmic softening for antiprotons, with the local processes and absolute normalizations kept distinct.

\begin{figure}[htbp]
\centering
\includegraphics[width=83.5mm]{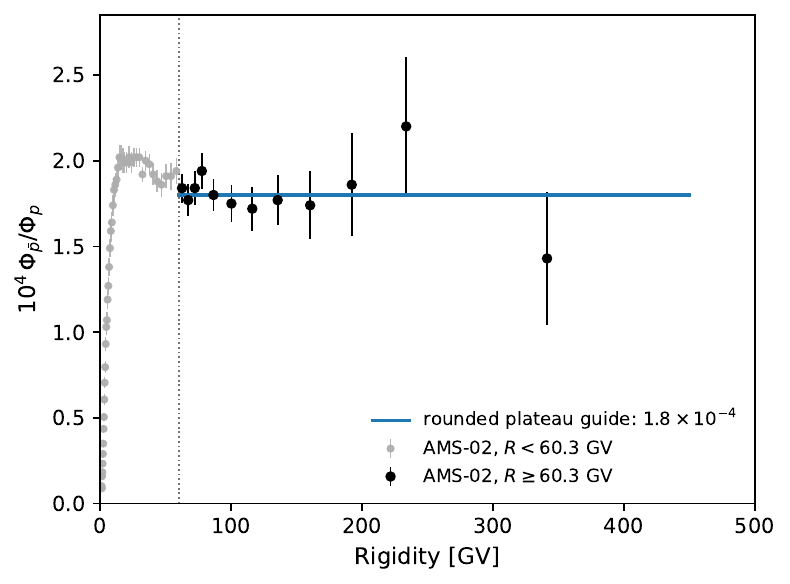}
\caption{AMS-02 antiproton-to-proton intensity ratio on a linear rigidity axis \cite{AMSbarp2016}. Gray points give low-rigidity context; black points have bin lower edges at or above 60.3 GV. The horizontal line marks the rounded observed plateau level $1.8\times10^{-4}$. It is a visual guide, not a fit or an absolute-normalization prediction. Error bars combine statistical and total systematic uncertainties in quadrature.}
\label{fig:pbar-ratio}
\end{figure}

Figure~\ref{fig:pbar-ratio} compares this leading behavior with the published ratio \cite{AMSbarp2016}. The horizontal guide at the observed level $1.8\times10^{-4}$ displays the nearly flat term; the small nonzero correction is estimated in Eq.~\eqref{eq:pbar-scaling}. The charge-resolved comparison thus tests the approximate index preservation associated with the common geometry.

The production-scaling conditions must cover the contributing parent rigidities. Cross-section evolution, proton curvature, and other nuclear channels modify $Q_{\bar p}/\Phi_p$ \cite{Winkler,Korsmeier}. Their contributions enter separately: the observed ratio slope is the source-to-proton slope minus $\delta_{\rm eff}$. These source effects can exceed the small response correction estimated here. Equation~\eqref{eq:pbar-geometric-softening} therefore specifies the response contribution, with detailed predictions for the full ratio also depending on ordinary production.

\section{Discussion and conclusion}
\label{sec:discussion}

Galactic cosmic rays provide the macroscopic histories examined in this work: their measured spectra contain both source information and accumulated environmental response \cite{StrongReview}. AMS-02 resolves these spectra by species and charge, allowing the cosmic-ray positron structure and the antiproton-to-proton regularity to be examined together \cite{AMSpositron,AMSelectron,AMSbarp2016}. In this setting, the time-orientation framework connects the two observations through a common history-overlap geometry, while retaining standard local QED and QCD and positive-energy detector records. Equation~\eqref{eq:response} captures the leading spectral form, and Eq.~\eqref{eq:pbar-geometric-softening} estimates the small residual antiproton softening.

For positrons, $q\simeq0.30$ gives $\xi_G=0.027$, moving the empirical $E^3$-weighted electron maximum from 8--10 GeV to the 300--400 GeV region. In the equal-injection illustration, particle-number conservation connects the relative height to energy redistribution and spatial dilution. For $\Gamma=2.5$, the observed order-$0.1$ weighted height corresponds to $a\simeq4$, which gives $0.095\simeq0.1$. The extended comparison retains these fixed scales. For antiprotons, ordinary secondary production retains the parent-proton history. Applying the same overlap estimate to logarithmic deformation gives $\delta_{\rm eff}\sim\xi_G\delta$, of order $10^{-2}$ for an ordinary softening of a few tenths. In the source-scaling limit, the ratio is therefore nearly flat with a small residual tilt.

Together, the two AMS channels provide a compelling empirical motivation to investigate time orientation as an organizer of antimatter histories. Within the geometric approximation, the same overlap scale connects the positron structure to the small antiproton slope correction. The physical connection concerns accumulated response across leptonic and hadronic observations, with each channel retaining its own production process.

The shared weighting is an effective geometric approximation, with representative history sampling and an order-one coefficient. The electron reference retains primary and heliospheric contributions; the positron broadening $a$ in the equal-injection illustration is inferred from the relative weighted height. The positron continuation has visible deviations, and the antiproton plateau height is not predicted. Production-scaling corrections enter separately. Conventional secondary models remain viable \cite{Boudaud,Heisig}, so the spectral correspondences do not uniquely identify the microscopic origin. These conditions define the scope of the history-overlap interpretation.

Further theory must derive the orientation-dependent weights and their common geometric reduction. Broader spectra and controlled positron experiments, with observables calculated for their own preparation and detection conditions, offer independent tests. AMS gives this foundational time-orientation interpretation concrete empirical direction through explicit spectral consequences.

\Needspace{8\baselineskip}
\section*{Acknowledgments}
This work was supported by Academia Sinica, the National Science and Technology Council of Taiwan, and National Cheng Kung University. The author used ChatGPT (OpenAI) for language editing, manuscript preparation, and analytical and code checks, reviewed the output, and takes full responsibility for the work.

\section*{Data availability}
The input spectra are the published AMS-02 supplemental tables \cite{AMSpositron,AMSelectron,AMSbarp2016}. The accompanying source package includes the input tables, code reproducing the fixed central curves and the propagated electron-reference errors, and numerical checks of particle-count conservation, production scaling, and the proposed geometric softening estimate. No spectral or amplitude optimization is used in the figures.

\section*{Conflict of interest}
The author declares no competing interests.

\begingroup
\small
\setlength{\bibsep}{2pt}

\endgroup

\begin{thebibliography}{99}

\bibitem{Stueckelberg}
E.~C.~G. Stueckelberg,
La signification du temps propre en m\'ecanique ondulatoire,
Helv. Phys. Acta \textbf{14} (1941) 322.

\bibitem{Feynman}
R.~P. Feynman,
The Theory of Positrons,
Phys. Rev. \textbf{76} (1949) 749.

\bibitem{Peskin}
M.~E. Peskin and D.~V. Schroeder,
\textit{An Introduction to Quantum Field Theory}
(Addison-Wesley, 1995).

\bibitem{YangTime}
Y.~Yang,
Time Orientation in Quantum Field Theory,
Zenodo (2026),
\href{https://doi.org/10.5281/zenodo.20821313}{doi:10.5281/zenodo.20821313}.

\bibitem{StrongReview}
A.~W. Strong, I.~V. Moskalenko, and V.~S. Ptuskin,
Cosmic-Ray Propagation and Interactions in the Galaxy,
Annu. Rev. Nucl. Part. Sci. \textbf{57} (2007) 285--327,
\href{https://doi.org/10.1146/annurev.nucl.57.090506.123011}{doi:10.1146/annurev.nucl.57.090506.123011},
arXiv:astro-ph/0701517.

\bibitem{AMSpositron}
M.~Aguilar \textit{et al.} (AMS Collaboration),
Towards Understanding the Origin of Cosmic-Ray Positrons,
Phys. Rev. Lett. \textbf{122} (2019) 041102.

\bibitem{AMSelectron}
M.~Aguilar \textit{et al.} (AMS Collaboration),
Towards Understanding the Origin of Cosmic-Ray Electrons,
Phys. Rev. Lett. \textbf{122} (2019) 101101.

\bibitem{AMSbarp2016}
M.~Aguilar \textit{et al.} (AMS Collaboration),
Antiproton Flux, Antiproton-to-Proton Flux Ratio, and Properties of Elementary Particle Fluxes in Primary Cosmic Rays Measured with the Alpha Magnetic Spectrometer on the International Space Station,
Phys. Rev. Lett. \textbf{117} (2016) 091103.

\bibitem{HooperPulsars}
D.~Hooper, P.~Blasi, and P.~D. Serpico,
Pulsars as the Sources of High Energy Cosmic Ray Positrons,
JCAP \textbf{01} (2009) 025.

\bibitem{YangYeung}
Y.~Yang and W.~B. Yeung,
Highly Suppressed Detection Probability of the Primordial Antimatter in the Present-Day Universe,
Chin. J. Phys. \textbf{77} (2022) 1544; arXiv:2604.02388.

\bibitem{YangPositron}
Y.~Yang, \emph{A Possible Advanced Positron Signal in AMS-02 Measurements}, arXiv:2604.25542 [hep-ph] (2026).

\bibitem{BlumenthalGould}
G.~R. Blumenthal and R.~J. Gould,
Bremsstrahlung, Synchrotron Radiation, and Compton Scattering of High-Energy Electrons Traversing Dilute Gases,
Rev. Mod. Phys. \textbf{42} (1970) 237.

\bibitem{StrongMoskalenko}
A.~W. Strong and I.~V. Moskalenko,
Propagation of Cosmic-Ray Nucleons in the Galaxy,
Astrophys. J. \textbf{509} (1998) 212--228,
\href{https://doi.org/10.1086/306470}{doi:10.1086/306470},
arXiv:astro-ph/9807150.

\bibitem{Evoli}
C.~Evoli, G.~Morlino, P.~Blasi, and R.~Aloisio,
AMS-02 Beryllium Data and Its Implication for Cosmic Ray Transport,
Phys. Rev. D \textbf{101} (2020) 023013,
\href{https://doi.org/10.1103/PhysRevD.101.023013}{doi:10.1103/PhysRevD.101.023013},
arXiv:1910.04113.

\bibitem{PDGCosmic}
J.~Alvarez-Mu\~niz, Z.~Cao, U.~F.~Katz, P.~Mertsch, and C.~Spiering,
Cosmic Rays, in \textit{Review of Particle Physics}, Particle Data Group,
revised March 2026,
\href{https://pdg.lbl.gov/2026/reviews/rpp2026-rev-cosmic-rays.pdf}{online review}.

\bibitem{Winkler}
M.~W. Winkler,
Cosmic Ray Antiprotons at High Energies,
JCAP \textbf{02} (2017) 048.

\bibitem{Korsmeier}
M.~Korsmeier, F.~Donato, and M.~Di Mauro,
Production Cross Sections of Cosmic Antiprotons in the Light of New Data from the NA61 and LHCb Experiments,
Phys. Rev. D \textbf{97} (2018) 103019.

\bibitem{Boudaud}
M.~Boudaud, Y.~G\'enolini, L.~Derome, J.~Lavalle, D.~Maurin, P.~Salati, and P.~D. Serpico,
AMS-02 Antiprotons' Consistency with a Secondary Astrophysical Origin,
Phys. Rev. Research \textbf{2} (2020) 023022.

\bibitem{Heisig}
J.~Heisig, M.~Korsmeier, and M.~W. Winkler,
Dark Matter or Correlated Errors: Systematics of the AMS-02 Antiproton Excess,
Phys. Rev. Research \textbf{2} (2020) 043017.
\end{thebibliography}
\end{document}